\documentclass[twocolumn,showpacs,preprintnumbers,amsmath,amssymb]{revtex4}
\usepackage{graphicx} 
\usepackage{dcolumn}  
\usepackage{latexsym}
\usepackage{color}

\def\DEL#1{{\textcolor{green}{}}}         

\begin{document}

\title{The Lorentz force effect on the On-Off dynamo intermittency }

\author{Alexandros Alexakis, Yannick Ponty}
\affiliation{Laboratoire Cassiop\'ee,
             Observatoire de la C\^ote d'Azur, \\
             BP 4229, Nice Cedex 04, France}

\begin{abstract}

An investigation of the dynamo instability close to the threshold 
produced by an ABC forced flow is presented. 
We focus on the on-off intermittency behavior of the
dynamo and the counter-effect of the Lorentz force in the non-linear 
stage of the dynamo. 
The Lorentz force drastically alters the statistics of the 
turbulent fluctuations of the flow and reduces their amplitude. 
As a result much longer burst (on-phases) 
are observed than what is expected based on the amplitude of
the fluctuations in the kinematic regime of the dynamo.
For large Reynolds numbers, the duration time of the ``On'' phase
follows a power law distribution, while for smaller Reynolds
numbers the Lorentz force completely
kills the noise and the system transits from a chaotic state into a
``laminar'' time periodic flow. 
The behavior of the On-Off
intermittency as the Reynolds number is increased is also examined.
The connections with dynamo experiments and theoretical modeling
are discussed.

\end{abstract}


\maketitle  

\section{Introduction}

Dynamo action, the self amplification of magnetic field due to the stretching
of magnetic field lines by a
flow, is considered to be the main mechanism for the generation of
magnetic fields in the universe \cite{mhdbooks}. To that respect
many experimental groups have successfully attempted to reproduce
dynamos in liquid sodium laboratory experiments
\cite{gailitis2000,gailitis2001,gailitis2004,muller2000,stieglitz2001,monchaux2007,berhanu2007}.
The induction experiments
\cite{odier1998,peffley2000a,peffley2000b,frick2002,bourgoin2002,nornberg2006a,nornberg2006b,stepanov2006,volk2006,bourgoin2006}
studying the response of an applied magnetic field inside a
turbulent metal liquid represent also a challenging science. With or
without dynamo instability the flow of a conducting fluid
 forms complex system, with a large degree of freedoms and a wide
branch of non linear behaviors.

In this work we focus on one special behavior: the On-Off
intermittency or blowout bifurcation \cite{pomeau1980,platt1993}.
 On-off intermittency 
is present in chaotic dynamical systems for which
there is an unstable invariant manifold in the phase space
such that the unstable solutions have a growth rate that varies strongly in time
taking both positive and negative values.
If the averaged growth rate is sufficiently smaller
than the fluctuations of the instantaneous growth rate, then the solution can
exhibit on-off intermittency where bursts of the amplitude of the distance
from the invariant manifold are observed (when the growth rate is positive)
followed by a decrease of the amplitude (when the growth rate is negative).
(See \cite{sweet2001a,sweet2001b} for a more precise definition).

On-Off intermittency has been observed in different
physical experiments including electronic devices,
electrohydrodynamic convection in nematics, gas discharge plasmas,
and spin-wave instabilities \cite{phys-onoff}.
In the MHD context, near the dynamo instability onset, 
the On-Off intermittency has been investigated
by modeling of the Bullard dynamo \cite{leprovost2006}.
Using direct numerical simulation \cite{sweet2001a,sweet2001b} were able to observe On-Off
intermittency solving the full MHD equations for the ABC dynamo,
(here we present an extended work of this particular case).
On-Off intermittency has also been found recently for a Taylor-Green flow \cite{ponty2007b}.
Finally, recent liquid metal experimental results (VKS) \cite{pinton2007}
show some intermittent behavior, with features reminiscent of on-off self-generation
that motivated our study.


For the MHD system we are investigating the evolution of the magnetic energy
$E_b=\frac{1}{2}\int {\bf b}^2 dx^3$ is given by
$\partial_t  { E_b} = \int {\bf b( \cdot b \nabla) u - \eta (\nabla b)^2} dx^3$.
If the velocity field has a chaotic behavior in time
the right hand side of the equation above can take positive or negative values
and can be modeled as multiplicative noise. A simple and proved very useful way
to model the behavior of the magnetic field during the on-off intermittency
is using a stochastic differential equation (SDE-model)
\cite{pomeau1980,platt1993,fujisaka,Yu1990,ott1004,Platt1994,Heagy1994,Venka1995,Venka1996,aumaitre2006,aumaitre2005}:

\begin{equation}
\partial_t E_b = (a+\xi) E_b - NL(E_b)
\label{SDE}
\end{equation}
where $ E_b$ is the magnetic energy,
      $a$ is the long time averaged growth rate,
      $\xi$  models the noise term typically assumed to be white
            (see however \cite{aumaitre2006,aumaitre2005}) and of amplitude $D$
             such that $\langle \xi(t)\xi(t') \rangle = 2D\delta(t-t')$.
            $NL$  is a non-linear term that guaranties the saturation
            of the magnetic energy to finite values typically taken to be
            $NL(X)=X^3$ for investigations of supercritical bifurcations
            or $NL(X)=X^5-X^3$ for investigations of subcritical bifurcations.
            Alternative, an upper no-flux boundary is imposed at $E_b=1$.
            In all these cases (independent of the non-linear saturation mechanism) 
            the above SDE leads to the stationery distribution function
            that for $0<a<D$
            has a singular behavior at $E_b=0$: $P(E_b)\sim E_b^{a/D-1}$ 
            indicating that the systems spends a 
            lot of time in the neighborhood of $E_b=0$. This is singularity
            is the signature of On-Off intermittency. 
            Among other predictions of the SDE model here we note that 
            the distribution of the duration time of the ``off" phases
            follows a power law behavior $PDF(\Delta T_{off})\sim \Delta T_{off}^{-1.5}$,
            all moments of the magnetic energy follow a linear scaling with
            $a$, $\langle E_b^m \rangle \sim a$, and for $a=0$ the set of the
            burst has a fractal dimension $d=1/2$ \cite{Platt1994,Heagy1994,Venka1995,Venka1996}.  

In this dynamical system eq.(\ref{SDE}) however the noise amplitude  
or the noise proprieties do not depend on the amplitude of the magnetic energy.
However, in the MHD system, when the non-linear regime is reached,
the Lorentz force has a clear effect on the the flow such as
the decrease of the small scale fluctuation, and the
decrease of the local Lyapunov exponent \cite{cattaneo1996,zienicke1998}.
Some cases, the flow is altered so strongly that the MHD dynamo system jumps into an other attractor,
that cannot not sustain any more the dynamo instability \cite{brummell}.
Although the exact mechanism of the saturation of the MHD dynamo is still an open question
that might not have a universal answer, it is clear that both the large scales and the turbulent fluctuations
are altered in the non-linear regime and need to be taken into account in a model.

\begin{figure}
\includegraphics[width=8cm]{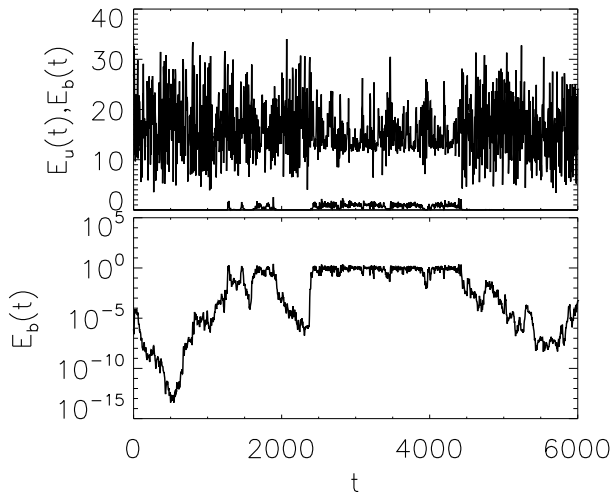}
\caption{A typical example of a burst.
The top panel shows the evolution of the kinetic energy (top line)
and magnetic energy (bottom line). The bottom panel shows the evolution
of the magnetic energy in a log-linear plot.
During the on phase of the dynamo the amplitude of the noise of the kinetic energy
fluctuations is significantly reduced.
The runs were for the parameters $Gr=39.06$ and $G_M=50.40$.
\label{fig1}}
\end{figure}

Figure \ref{fig1} demonstrates this point, by showing the evolution of the kinetic and
magnetic energy as the dynamo goes through On- and Off- phases.
During the On phases although the magnetic field energy is an order of magnitude smaller
than the kinetic energy 
both the mean value and
the amplitude of the observed fluctuations of the kinetic energy
are significantly reduced. 
As a result the On-phases last a lot longer
than what the SDE-model would predict.
With our numerical simulations, we aim to describe which of the On-Off intermittency proprieties are affected
through the Lorentz force feed-back.

This paper is structured as follows. In the next section \ref{Nmethod} we discuss the numerical method used.
In section \ref{Onset} we present the table of our numerical runs and discuss the dynamo onset. 
Results for
small Reynolds numbers investigating the transition from a laminar dynamo to on-off intermittency
are presented in \ref{Route}, and the results on fully developed on-off intermittency behavior are given in
section \ref{Onoff}. Conclusions, and implications on modeling and on  the laboratory experiments are
given in the last section \ref{Cons}.

\section{Numerical method}
\label{Nmethod}
Our investigation is based on the numerical integration of the
classical incompressible MagnetoHydroDynamic equations (MHD) (\ref{eq:MHD}) in a
full three dimensional periodic box of size $2\pi$, with a parallel pseudo-spectral code.
The MHD equations are:
\begin{eqnarray}
\partial_t {\bf u} + {\bf u} \cdot \nabla {\bf u} &=& -\nabla P
+ (\nabla \times {\bf b}) \times  {\bf b} + \nu  \nabla^2 {\bf u}  +{\bf f} \nonumber \\
\partial_t {\bf b}  &=& \nabla \times ({\bf u} \times {\bf b}) \cdot {\bf u}
+ \eta \nabla^2 {\bf b}
\label{eq:MHD}
\end{eqnarray}
along with the divergence free constrains $\nabla \cdot {\bf u} = \nabla \cdot {\bf b} = 0$.
Where {$\bf u $} is the velocity,
${\bf b}$ is  the magnetic field (in units of Alfv\'en velocity),
$\nu$  the molecular viscosity and $\eta$ the magnetic diffusivity.
${\bf f}$ is an externally applied force that in the
current investigation is chosen to be the ABC forcing \cite{arnold1965} explicitly given by
\begin{eqnarray}
{\bf f} = \hat{x} ( A \sin(k_zz) + C\cos(k_yy) )\nonumber \\
          \hat{y} ( B \sin(k_xx) + A\cos(k_zz) )\\
          \hat{z} ( C \sin(k_yy) + B\cos(k_xx) ) \nonumber
\label{eq:forcing}
\end{eqnarray}
with all the free parameters chosen to be unity $A=B=C=k_x=k_y=k_z=1$.

The MHD equations have two independent control parameters that are generally
chosen to be the kinetic and magnetic Reynolds numbers defined by:
$Re=UL/\nu$ and $R_M=UL/\eta$ respectably, where $U$ is chosen to be the root mean square of the velocity
(defined by $U=\sqrt{2 E_u/3}$, where $E_u$ is the total kinetic energy of the velocity)
and $L$ is the typical large scale here taken $L=1.0$.
Alternatively we can use the amplitude of the forcing to parametrize our system
in which case we obtain the kinetic and magnetic Grashof numbers
$Gr=FL^3/\nu^2$ and $G_M=FL^3/\nu\eta$ respectably. Here $F$ is the amplitude of the force
that is taken to be unity 
$F=\sqrt{(A^2+B^2+C^2)/3}=1$ following the notation of \cite{podvigina1994}.

We note that in the laminar limit the two different sets of control parameters are identical
$Gr = Re$ and $G_M = R_M$ but in the turbulent regime the scaling $Gr \sim Re^2$ and $G_M \sim ReR_M$
is expected. In the examined parameter range the velocity field fluctuates in time generating uncertainties
in the estimation of the root mean square of the velocity and then the Reynolds numbers.
For this reason in this work we are going to use the Grashof numbers
as the control parameters of our system.

Starting with a statistically saturated velocity,
we investigate the behavior of the kinetic and magnetic energy in time by introducing
a small magnetic seed at $t=0$ and letting the system evolve. When the
magnetic Grashof (Reynolds) number is sufficiently large the magnetic energy
grows exponentially in time reaching the dynamo instability.
We have computed the dynamo onset for different kinematic Grashof (Reynolds) numbers (III.A)
starting from small $Gr=11.11$ that the flow exhibits laminar ABC behavior
to larger values of $Gr$ (up to $Gr=625.0$) that the flow is relatively turbulent.

Typical duration of the runs
were $10^5$ turn over times although in some cases even much longer integration time
was used.
For each run during the kinematic stage of the dynamo the
finite time growth rate
$a_\tau(t)=\tau^{-1}\log(E_b(t+\tau)/E_b(t)$ was measured.
The long time averaged growth rate was then determined as
$a=\lim_{\tau\to \infty} a_\tau(0)$ and
the amplitude of the noise $D$ was the measured based on
$D=\tau \langle (a-a_\tau)^2 \rangle/2$ (see \cite{sweet2001a,sweet2001b}).
Typical value of $\tau$ was 100 while the for long time average
the typical averaging time ranged from $10^4$ to $10^5$ depending on the run.
The need for long computational time in order to obtain good statistics
restricted our simulations to low resolutions that varied from $32^3$
(for $Gr \le 40.0$) to  $64^3$ (for $Gr > 40$).

\section{Numerical results}

\subsection{Dynamo onset}
\label{Onset}

The ABC flow
is a strongly helical Beltrami flow with chaotic Lagrangian trajectories \cite{dombre1986}.
The kinematic dynamo instability of the ABC flow, even with one of the amplitude coefficients set to zero
($2D^{1/2}$ flow)
\cite{galloway1992,ponty1995} has been study intensively
\cite{arnold1983,galloway1986,teyssier2006,galanti1992,archontis2003},
especially for fast dynamo investigation
\cite{childress1995,moffat1985,bayly1988,finn1988a,finn1988b}.
In the laminar regime and for the examined case
where all the parameter of the ABC flow are equal to the unity
(equations (\ref{eq:forcing}),
the flow has dynamo in the range $8.9 \lesssim G_M \lesssim 17.8$ and $24.8<G_M$ \cite{arnold1983,galloway1986}.
In this range the magnetic field is growing near the stagnation point of the flow,
producing ``cigar'' shape structures  aligned along the unstable manifold.

As the kinematic Grashof number is increased,
a critical value is reached ($ Gr = Re \sim 13. $) that
the hydrodynamic system becomes unstable.
After the first bifurcation, further increase of the kinematic Grashof (Reynolds) number,
leads the system to jump to different attractors \cite{podvigina1994,podvigina1999},
until finally the fully turbulent regime is reached.

The On-off intermittency dynamo was studied with the forcing ABC
by \cite{sweet2001a,sweet2001b} although their study was focused on a single value
of the mechanical Grashof number while the magnetic Grashof number was varied.
We expand this work by varying both parameters.
For each kinematic Grashof number a set of numerical runs
were performed varying the magnetic Grashof number.
A table of the different Grashof (Reynolds) numbers examined is shown in table I.
The case examined in \cite{sweet2001a,sweet2001b} is closer to the set of runs with $Gr=39.06$
although here examined at higher resolution. 

\begin{figure}
\includegraphics[width=8cm]{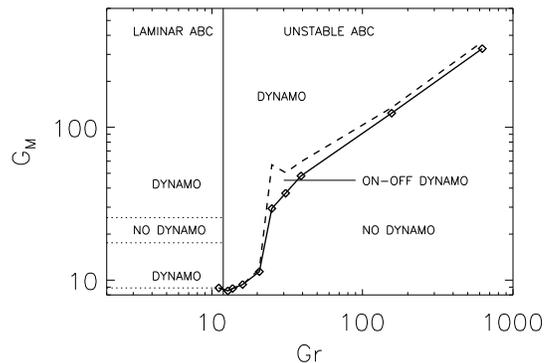}
\caption{Critical magnetic Grashof number $G_{Mc}$ that the dynamo instability
is observed (solid line) and the critical magnetic Grashof number $G_{Mo}$
where the on-off intermittency is disappears (dashed line).
\label{fig2}}
\end{figure}

First, we discuss the dynamo onset. For each kinetic Grashof number
the critical magnetic Grashof number $G_{Mc}$ is found
and recorded in table I.
For our lowest kinematic Grashof number $Gr=11.11$ which corresponds
to a slightly smaller value than the critical value that hydrodynamic instabilities are present,
the flow is laminar and the two windows of dynamo instability \cite{arnold1983,galloway1986}
are rediscovered, shown in fig. \ref{fig2}.
At higher Grashof number, the hydrodynamic system is not stable anymore,
and the two dynamo window mode disappear,
to collapse in only one (see fig.\ref{fig2}).
The critical magnetic Reynolds number
is increasing with the Grashof (Reynolds) number fig.\ref{fig2},
and saturates
at very large values of $Gr$ \cite{mininni} that are far beyond the
range examined in this work.

\begin{table}
\caption{\label{table:runs} Parameters used in the simulations.
$G_{Mc}$ is the critical magnetic Grashof
that the dynamo instability begins and $G_{Mo}$ is the
critical magnetic Grashof that the dynamo instability stops
having on-off behavior. Thus, on-off intermittency is observed
in the range $G_{Mc}< G_{M}<G_{Mo}$.}
\begin{ruledtabular}
\begin{tabular}{|c||c|c|c|c|c|}
Run    & $\nu$  &    $Gr$    &   $Re$ &  $G_{Mc}$          &  $G_{Mo} $   \\
\hline
I      & 0.30  &   11.11     & 11.11        &   8.89--17.8, 24.0       & 8.89    \\
II     & 0.28  &   12.75     & 12.75/11.22  &   8.50       & 8.50    \\
III    & 0.25  &   16.00     & 14.82        &   9.35       & 9.35    \\
IV     & 0.22  &   20.66     & 11.65        &   11.3       & 11.8    \\
VI     & 0.20  &   25.00     & 18.45        &   29.4       & 56.8    \\
VII    & 0.18  &   30.86     & 19.47        &   37.0       & 50.5    \\
VII    & 0.16  &   39.06     & 20.60        &   48.0       & 59.5    \\
VII    & 0.08  &  156.25     & 34.08        &   123.7      & 137.    \\
VII    & 0.04  &  625.00     & 67.20        &   327.2      & 362.    \\
\end{tabular}
\end{ruledtabular}
\end{table}

\subsection{Route to the On-Off intermittency}
\label{Route}

\begin{figure}
\includegraphics[width=8cm]{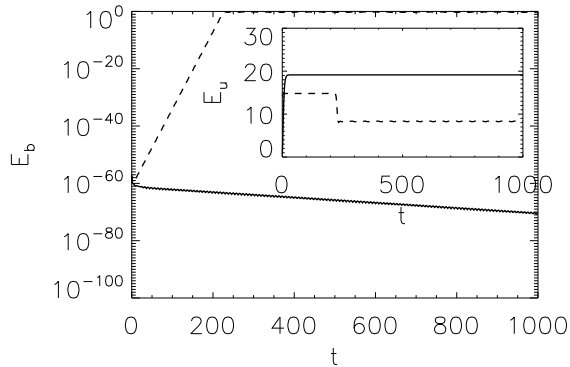}
\caption{Kinetic (inset) and magnetic energy for the run with
$Gr=12.75$ and $G_M=22.32$ for two runs starting with different
initial conditions for the velocity field.
The first flow (solid line) is attracted to the laminar ABC flow and gives no dynamo
the second flow (dashed line) is attracted to a new solution that gives dynamo.
\label{fig3}}
\end{figure}

The first examined Grashof number beyond the laminar regime is $Gr=12.75$ (run II).
In this case two stable solutions of the Navier-Stokes co-exist. 
Depending of the initial starting condition, this hydrodynamic system converges into one of the two attractors.   
The two velocity fields have
different critical magnetic Grashof numbers. 
The first solution is the laminar flow that shares the same dynamo properties
with the smaller Grashof number flows.
For the second flow however 
the previous stable window between $G_M\simeq 17.8$ and $G_M=24.0$
disappears and the critical magnetic Grashof number
now becomes $G_{Mc}=8.50$, resulting in only one instability window. 
Figure \ref{fig3} demonstrates the different dynamo properties of the two solutions.
The evolution of the kinetic and magnetic energy of two runs is shown with the same parameters $Gr,G_M$
but with different initial conditions for the velocity field.
$G_M$ is chosen in the range of the no-dynamo window of the laminar ABC flow.

This choice of $Gr$ although it exhibits interesting behavior does not give
on-off intermittency since both hydrodynamic solutions are stable in time.
The next examined Grashof number (III), gives a chaotic behavior of the hydrodynamic
flow and accordingly a ``noisy" exponential growth rate for the magnetic field.
\begin{figure}
\includegraphics[width=8cm]{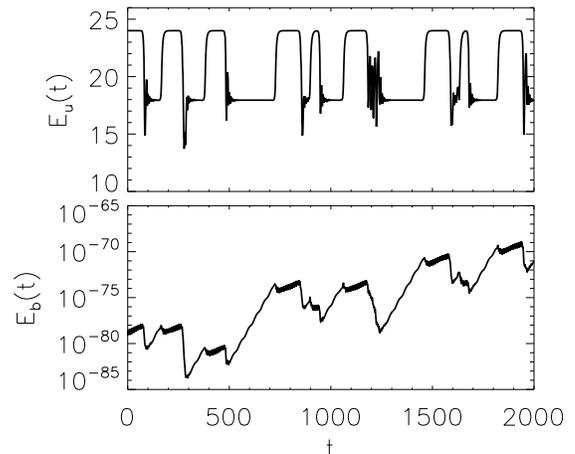}
\caption{The evolution of the kinetic (top panel) and magnetic (bottom panel) energy
for the run with $Gr=16.0$ and $G_M=9.39$.
\label{fig4}}
\end{figure}
The evolution of the kinetic energy and the magnetic energy in the kinematic regime is
shown in fig. \ref{fig4} for a relatively short time interval.
The kinetic energy ``jumps" between the values of the kinetic energy of the two states 
that were observed
to be stable at smaller Grashof numbers in a chaotic manner.
Accordingly the magnetic energy grows or decays depending on the state of the hydrodynamic flow,
in a way that very much resembles a biased random walk in the log-linear plane.
Thus, this flow is expected to be a good candidate for
on-off intermittency that could be modeled by the SDE model equations given in eq.\ref{SDE}.
However this flow did not result in on-off intermittency for all examined magnetic Grashof
numbers, even for the runs that the measured growth rate and amplitude of the noise were found
to satisfy the criterion $a/D<1$ for the existence of on-off intermittency.
What is found instead is that at the linear stage the the magnetic field grows in
a ``random" way but in the nonlinear stage the solution is ``trapped" in a stable periodic
solution and remains there throughout the integration time.
This behavior is demonstrated in fig. \ref{fig5} where the evolution of the  magnetic energy
is shown both in the linear and in the non-linear regime.
\begin{figure}
\includegraphics[width=8cm]{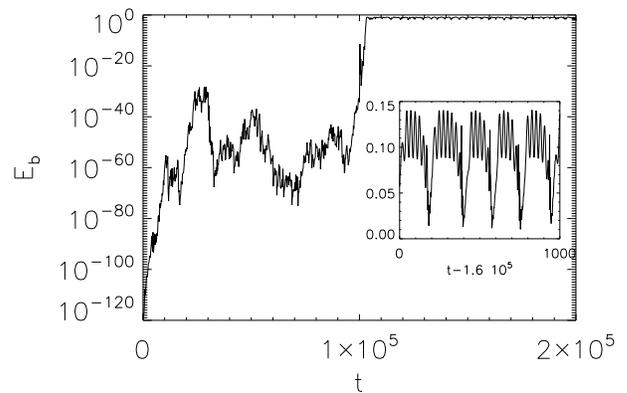}
\caption{The evolution of the magnetic energy
for the run with $Gr=16.0$ and $G_M=9.39$. At the linear stage
the logarithm of the magnetic energy grows like a random walk.
At the nonlinear stage however the solution is trapped in a stable
time periodic solution. The inset shows the evolution of the magnetic
energy in the nonlinear stage in a much shorter time interval.
The examined run has $a/D=0.022<1$.
\label{fig5}}
\end{figure}
`

\begin{figure}
\includegraphics[width=8cm]{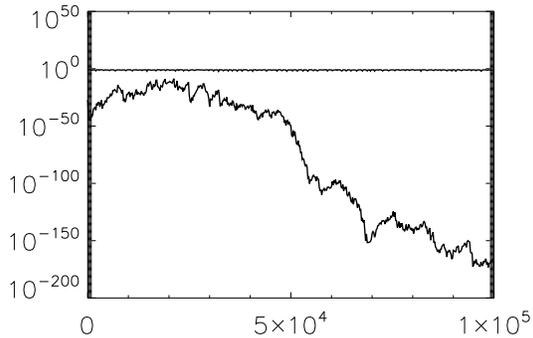}
\caption{Subcritical behavior of the ABC dynamo.
The evolution of the magnetic energy
for two runs with $Gr=16.0$ and $G_M=9.30$ starting with small amplitude magnetic
field (bottom line) and starting with an amplitude of the magnetic field at
the nonlinear stage (top almost straight line).
\label{fig6}}
\end{figure}

An other interesting feature of this flow is that exhibits subcriticality \cite{ponty2007b}.
The periodic solution that the dynamo simulations converged to in the nonlinear stage
appears to be stable even for the range of $G_M$ that no dynamo exists.
Figure \ref{fig6} shows the time evolution of two runs with the same parameters
$Gr,G_M$ one starting with very small amplitude of the magnetic field and one
starting using the output from one of the successful dynamo runs in the nonlinear stage.
Although the magnetic energy of the first run decays with time the nonlinear
solution appears to be stable.

\begin{figure}
\includegraphics[width=8cm]{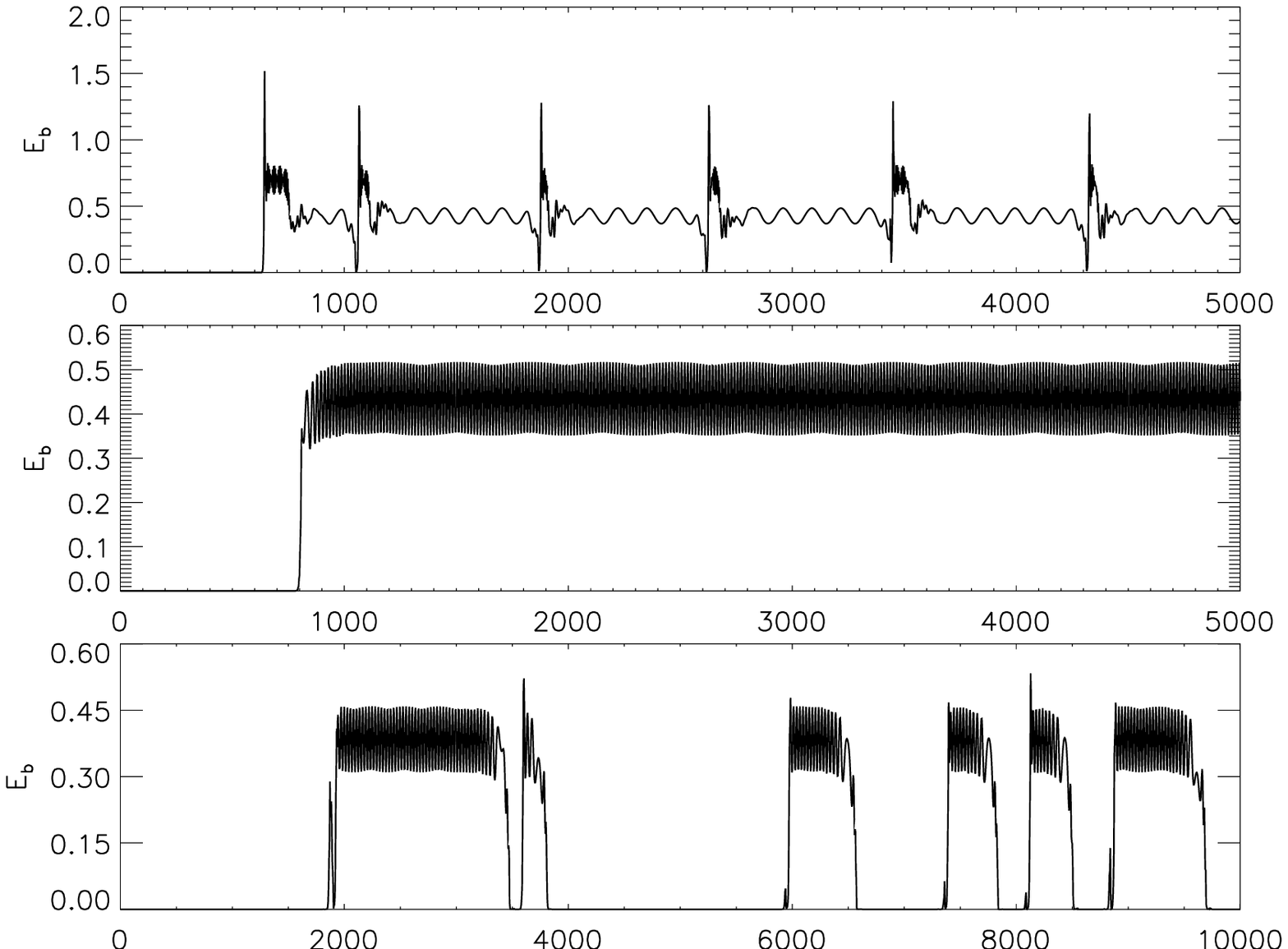}
\caption{Evolution of the magnetic energy for $Gr=20.66$ and
$G_M=20.66$ (top panel), $G_M=12.0$ (middle panel), $G_M=11.6$ (bottom panel).
\label{fig7}}
\end{figure}

The next examined Grashof number $G=20.66$ (IV) appears to be a transitory state
between the previous example and on-off intermittency that is examined in the next section.
Figure \ref{fig7} shows the evolution of the magnetic energy for three different
values of $G_M=20.66,12.0,11.6$ for all off which the ratio $a/D$ was measured and was found to be smaller than
unity and therefor are expected to give on-off intermittency based on the SDE model.
Only the bottom panel however (which corresponds to the value of $G_M=11.6$ closest
to the onset value $G_M=11.3$)  shows on-off intermittency.
A singular power law behavior of the pdf of the magnetic energy during the off phases (small $E_b$)
for the last run was observed to be in good agreement with the predictions of the SDE.
This is expected since for small $E_b$ the Lorentz force that is responsible for trapping the solution
in the nonlinear stage does not play any role.

\subsection{On-off intermittency}
\label{Onoff}

\begin{figure}
\includegraphics[width=8cm]{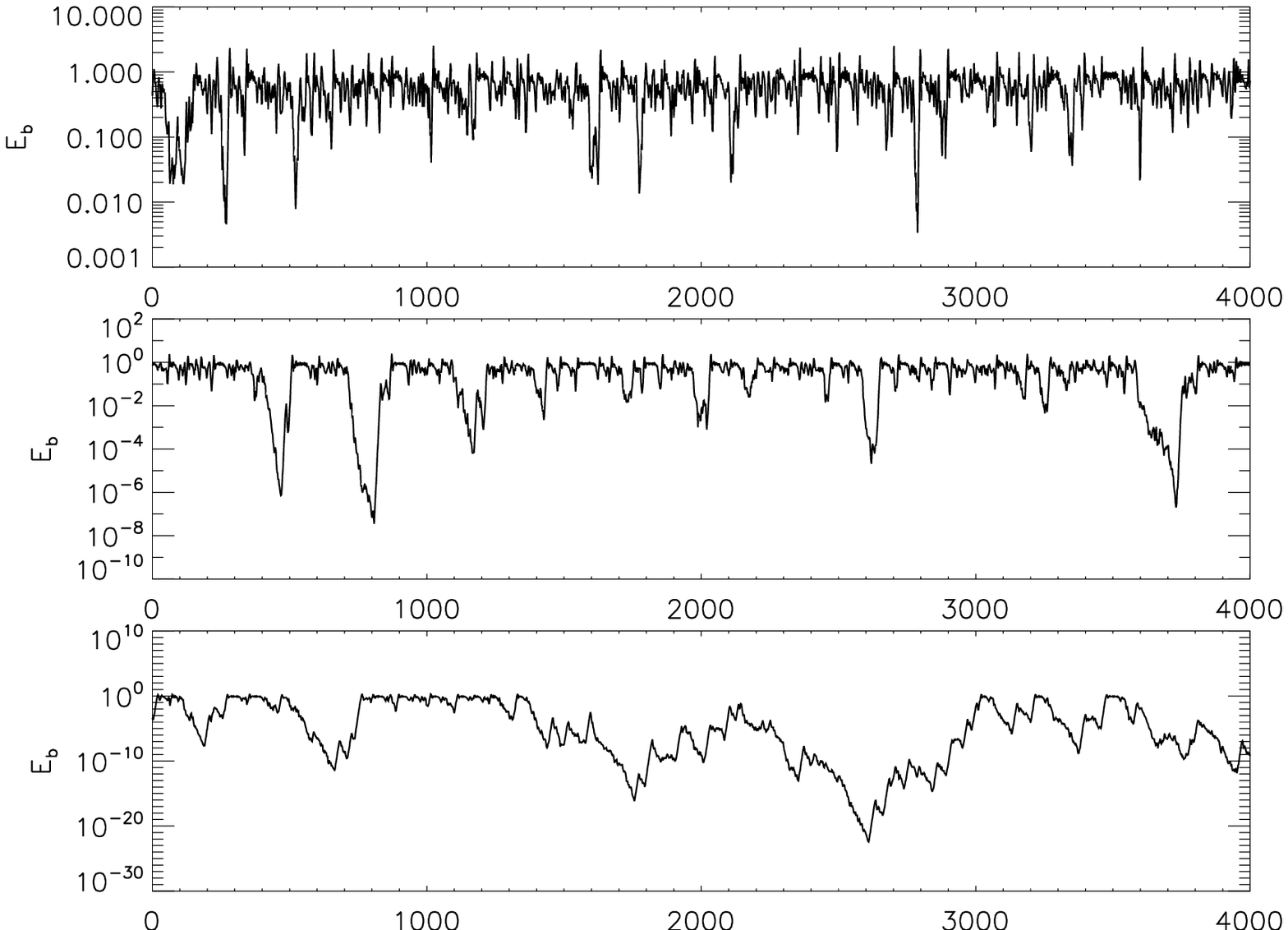}
\caption{Evolution of the magnetic energy for $Gr=20.66$ and
$G_M=20.66$ (top panel), $G_M=12.0$ (middle panel), $G_M=11.6$ (bottom panel).
\label{fig8}}
\end{figure}

\begin{figure}
\includegraphics[width=8cm]{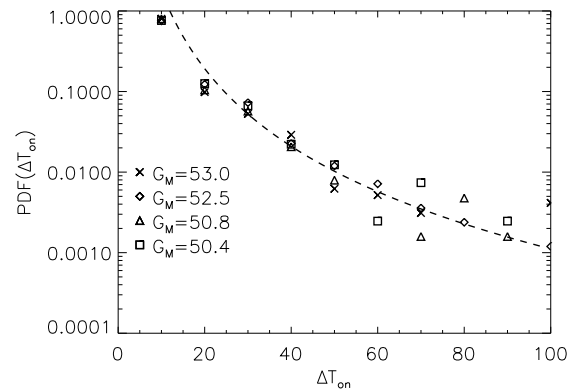}
\caption{Distribution of the ``on" times for the $Gr=39.06$ case and three different
values of $G_M$. The fit (dashed line) corresponds to the power-low behavior $\Delta T^{-3.2}$.
Here "on" time is considered the time that dynamo has magnetic energy $E_b>0.2$.
\label{fig9}}
\end{figure}

All the larger Grashof numbers examined display on-off intermittency
and there is no trapping of the solutions in the ``on" phase.
Figure \ref{fig8} shows an example of the on-off behavior for
$Gr=25.0$ and three different values of $G_M$
( $G_M=41.6$ (top panel), $G_M=35.7$ (middle panel), $G_M=31.2$ (bottom panel)).
As the critical value of $G_M$ is approached the ``on" phases of the dynamo (bursts)
become more and more rare as the SDE model predicts. Note however that
the ``on" phases of the dynamo last considerably long.
In fact the distribution of the duration of 
the ``on" phase $\Delta T_{on}$ is fitted best to a power law distribution rather than
an exponential that a random walk model with an upper no-flux boundary
would predict, as can be seen in fig. \ref{fig9}.

\begin{figure}
\includegraphics[width=8cm]{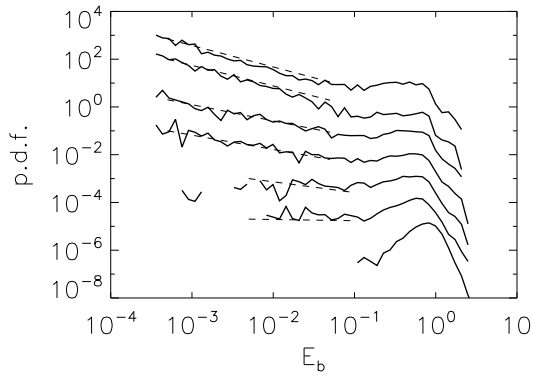}
\caption{The probability distribution functions of $E_b$, for $Gr=25$ and
and seven different values of $G_M$ (starting from the top line:
$G_M=31.2$,
$G_M=33.3$,
$G_M=35.7$,
$G_M=38.4$,
$G_M=41.6$,
$G_M=50.0$,
$G_M=83.3$.
The last case $G_M=83.3$ shows no on-off intermittency.
The dashed lines shows the prediction of the SDE model.
The pdfs have not been normalized for reasons of clarity.
\label{fig10}}
\end{figure}

The effect of the long duration of the ``on" times can also be seen in the pdfs of the
magnetic energy.
The pdfs for the $Gr=20.66$ for the examined $Gr$ are shown in figure
\ref{fig10}. For values of $G_M$ much larger from the critical value $G_{Mc}$
the pdf of the amplitude of the magnetic field
is concentrated at large values $E_b\simeq1$ producing a peak in the pdf curves. 
As $G_M$ is decreased approaching $G_{Mc}$ from above a singular behavior of
the pdf appears with the pdf having a power law behavior $\sim E_b^{-\gamma}$ for small $E_b$.
The closer the $G_M$ is to the critical value the singularity becomes stronger.
The dashed lines show the prediction of the SDE model $\gamma=1-a/D$.
The fit is very good for small $E_b$, however the SDE for a supercritical bifurcation fails to reproduce the
peak of the pdf at large $E_b$, that is due to the long duration of the ``on" phases.

\begin{figure}
\includegraphics[width=8cm]{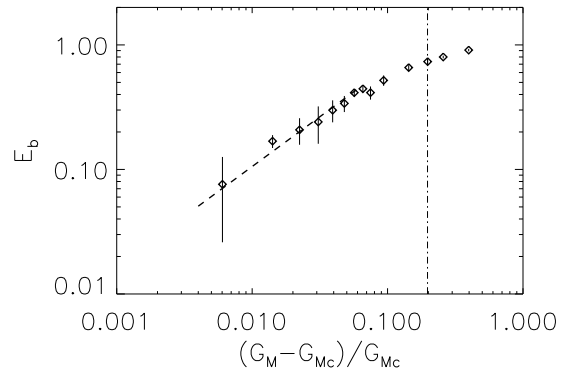}
\caption{Averaged magnetic energy as function of the relative deviation from
the critical magnetic Grashof number. The dash-dot vertical line indicates
the location of $(G_{Mo}-G_{Mc})/G_M$ beyond which On-Off intermittency is no longer present.  
\label{fig11}}
\end{figure}

An other prediction of the SDE model is that all the moments of the magnetic
energy $\langle E_b^m\rangle = \int PDF(E_b)E_b^m dE_b$ have a linear scaling
with the deviation of $G_M$ from the critical value $G_{Mc}$ provided that
the difference $G_M-G_{Mc}$ is sufficiently small.
This result is based on the assumption the singular behavior close to $E_b=0$
gives the dominant contribution to the pdf that is always true provided that
the ratio $a/D$ is sufficiently small. However if the system spends long times
in the ``on" phase the range of validity of the linear scaling of $\langle E_b \rangle $ with
$a \sim G_M-G_{Mc}$ is restricted to very small values of the difference $G_M-G_{Mc}$.
Figure \ref{fig11} shows the time averaged magnetic energy $\langle E_b \rangle $ as a function
of the relative difference $(G_M-G_{Mc})/G_M$ in a log-log scale. The dependence of $\langle E_b \rangle $
on the deviation of $G_M$ from the critical value appears to approach the linear scaling 
albeit very slow. The best fit from the six smallest values of $G_M$ 
shown in the fig.\ref{fig11} gave an exponent of 0.8 
(e.g $\langle E_b \rangle \sim (G_M-G_{Mc})^{0.8}$).
The small difference from the linear scaling ($\langle E_b \rangle \sim (G_M-G_{Mc})^{1}$ )
is probably because not sufficiently small 
deviations $(G_M-G_{Mc})$ were examined. 
We note however that there is a strong deviation from the linear scaling 
for values of $G_M$ close to $G_{Mo}$.

\begin{figure}[hb!]
\includegraphics[width=8cm]{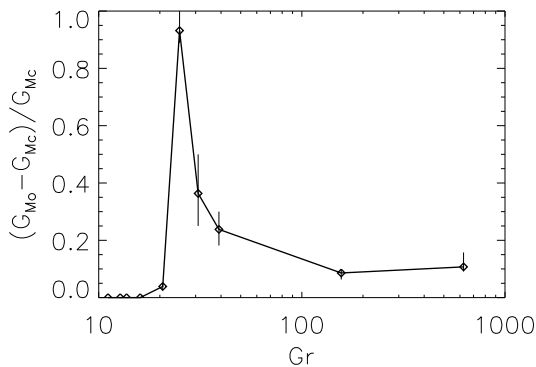}
\caption{The ratio $(G_{Mo}-G_{Mc})/G_{Mc}$ as a function of $G_r$ that
expresses the relative range that on-off intermittency is observed.
Error-bars correspond
to the smallest examined values of $G_M$ that no on-off intermittency was observed (upper error bar)
and the largest examined values of $G_M$ that    on-off intermittency was observed (lower error bar).
\label{fig12}}
\end{figure}

Of particular interest to the experiments is how the range of intermittency changes as
$Gr$ is increased. Typical $Gr$ numbers for the experiments are of the order of $Gr\sim Re^2 \sim 10^{12}$
that is not currently possible to be obtained in numerical simulations.
In figure \ref{fig2} we showed the critical magnetic Grashof number $G_{Mc}$ that dynamo instability
is observed and the critical magnetic Grashof number $G_{Mo}$ that the on-off intermittency is present.
$G_{Mc}$ was estimated by interpolation between the run with the smallest positive growth rate and
and the run with the smallest (in absolute value) negative growth rate.
The on-off intermittency range was based on the pdfs of the magnetic energy. Runs that the pdf had
singular behavior at $E_b\simeq 0$ are considered on-off while runs with smooth behavior at $E_b\simeq 0$
are not considered to show on-off intermittency. 
The slope of the pdfs (in log-log scale) for small $E_b$ were calculated and
the transition point $G_{Mo}$ was determined by
interpolation of the two slopes (see for example the bottom two curves in fig. \ref{fig10}).
In figure \ref{fig12} we show the ratio $(G_{Mo}-G_{Mc})/G_{Mc}$ as a function of $G_r$ that
expresses the relative range that on-off intermittency is observed. The error-bars correspond
to the smallest examined values of $G_M$ that no on-off intermittency was observed (upper error bar)
and the largest examined values of $G_M$ that    on-off intermittency was observed (lower error bar).
The range of on-off intermittency is decreasing as $Gr$ is increased probably reaching an asymptotic value.
However to clearly determine the asymptotic behavior of $G_{Mo}$ with $Gr$
would require higher resolutions that
the long duration of these runs does not allow us to perform.

\section{Discussion}
\label{Cons}

In this work we have examined how the on-off intermittency behavior
of a near criticality dynamo is changed as the kinematic Reynolds is
varied, and what is the effect of the Lorentz force in the
non-linear stage of the dynamo.
The predictions of \cite{Platt1994,Heagy1994,Venka1995,Venka1996},
linear scaling of the averaged magnetic energy with the deviation of the control parameter from its critical value,
fractal dimensions of the bursts,
distribution of the ``off" time intervals, and
singular behavior of the pdf of the magnetic energy      
that were tested numerically in 
\cite{sweet2001a,sweet2001b} were verified for a larger range of Kinematic Grashof numbers
when On-Off intermittency was present.
Note however that all these predictions are based on the statistics
of the flow in the kinematic stage of the dynamo.
However it was found that the Lorentz force can drastically alter the On-Off
behavior of the dynamo in the non-linear stage by quenching the noise.
For small Grashof numbers the Lorentz force can trap the
original chaotic system in the linear regime in to a time periodic state
resulting to no On-Off intermittency.
At larger Grashof numbers $Gr>20$ On-Off intermittency was observed
but with long durations of the ``on" phases that have a power law
distribution. These long ``on" phases result in a pdf that 
peaks at finite values of $E_b$. This peak can be attributed to
the presence of a subcritical instability or to the quenching of 
the hydrodynamic ``noise" at the nonlinear stage  
or possibly a combination of the two.
In principle the SDE model (eq.\ref{SDE}) can be modified to include these two effects:
a non-linear term that allows for a subcritical bifurcation and a $E_b$
dependent amplitude of the noise. There many possibilities to model
the quenching of the noise, however the nonlinear behavior might not
have a universal behavior and we do not attempt to suggest a specific model.

The relative range of the On-Off intermittency was found to decrease as the Reynolds number
was increased possibly reaching an asymptotic regime. However the limited number of Reynolds
numbers examined did not allow us to have a definite prediction for this asymptotic regime.
This question is of particular interest to the dynamo experiments 
\cite{gailitis2000,gailitis2001,gailitis2004,muller2000,stieglitz2001,monchaux2007,berhanu2007}
that until very recently \cite{pinton2007} have not detected On-Off intermittency .
There are many reasons that could explain the absence of detectable On-Off intermittency
in the experimental setups,
like the strong constrains imposed on the flow \cite{gailitis2004,muller2000}
that do not allow the development of large scale fluctuations
or the Earths magnetic field that imposes a lower threshold for the amplitude of the magnetic energy.
Numerical investigations at higher resolution and a larger variety of flows or forcing 
would be useful at this point to obtain a better understanding.

\begin{acknowledgments}
We thank F. P\'etr\'elis, J-F Pinton for fruitful discussions.
AA acknowledge the financial support from the ``bourse Poincar\'e''
of the  Observatoire de la C\^ote d'Azur and the Rotary Club district 1730.
YP thank CNRS Dynamo GdR, INSU/PNST, and INSU/PCMI Programs. Computer time was provided by
IDRIS, and the Mesocentre SIGAMM machine, hosted by Observatoire de la Cote d'Azur.
\end{acknowledgments}


\newpage

\begin{thebibliography}{99}

\bibitem{mhdbooks}
  H. K. Moffatt, Magnetic Field Generation in Electrically Conducting Fluids
 {\it Cambridge University Press, Cambridge}, (1978) ;
 F. Krause and K.-H.Radler, Mean-Field Magnetohydrodynamics and Dynamo Theory {\it Pergamon, Oxford,} (1980) ;
 E. N. Parker, Cosmical Magnetic Fields {\it Clarendon, Oxford, 1979} .

\bibitem{gailitis2000}
A. Gailitis, et al.,
{\it Phys. Rev. Lett.} {\bf 84}, 4365 (2000).

\bibitem{gailitis2001}
A. Gailitis, et al.,
{\it Phys. Rev. Lett.} {\bf 86}, 3024 (2001).


\bibitem{gailitis2004}
A. Gailitis, et al.,
{\it Physics of Plasmas}  {\bf 11} Issue 5,  2838--2843 (2004).


\bibitem{muller2000} U. Muller and R. Stieglitz, {\it Naturwissenschaften} {\bf 87}, 381 (2000).

\bibitem{stieglitz2001}  R. Stieglitz and U. M¨uller,
{\it Phys. Fluids}  {\bf 13},  561 (2001).

\bibitem{monchaux2007}
 R. Monchaux et al.,
{\it Phys. Rev. Lett.} {\bf 98}, 044502 (2007).

\bibitem{berhanu2007}
 M. Berhanu et al.,
{\it Europhys. Lett.} {\bf 77}, 59001 (2007).



\bibitem{odier1998}
P. Odier, J.-F. Pinton, S. Fauve
{\it Phys. Rev. E} {\bf 58}, 7397--7401 (1998).

\bibitem{peffley2000a}
N. L. Peffley, A. B. Cawthorne, and D. P. Lathrop,
{\it Phys. Rev. E} {\bf 61} 5287 (2000).


\bibitem{peffley2000b}
N. L. Peffley et al
{\it Geoph. J. Int.} {\bf 142}, 52--58 (2000).

\bibitem{frick2002}
P. Frick et al,
{\it Magnetohydrodynamics} {\bf 38} No. 1/2, 143--162, (2002).

\bibitem{bourgoin2002}
M. Bourgoin et al
{\it Phys. Fluids} {\bf 14}, 3046 (2002).

\bibitem{nornberg2006a}
M. D. Nornberg, E. J. Spence, R. D. Kendrick, C. M. Jacobson, and C. B. Forest,
{\it Phys. Rev. Lett.} {\bf 97}, 044503 (2006).

\bibitem{nornberg2006b}
M. D. Nornberg et al
{\it Phys. Plasmas } {\bf 13}, 055901 (2006).

\bibitem{stepanov2006}
R. Stepanov, R. Volk, S. Denisov, P. Frick, V. Noskov, and J.-F. Pinton
{\it Phys. Rev. E} {\bf 73}, 046310 (2006).

\bibitem{volk2006}
R. Volk, P. Odier, and J-F Pinton,
{\it Phys. Fluids } {bf 18}  085105 (2006).

\bibitem{bourgoin2006}
M. Bourgoin et al
{\it New Journal of Physics} {\bf 8},  329 (2006).



\bibitem{pomeau1980}
Y. Pomeau and P. Manneville,
{\it Commun. Math. Phys.} {\bf 74}, 1889 (1980).

\bibitem{platt1993}
 N. Platt, E. A. Spiegel and C. Tresser,
{\it Phys. Rev. Lett.}  {\bf 70} (3), 279--282 (1993).

\bibitem{sweet2001a} D. Sweet, E. Ott, J. M. Finn, T. M. Antonsen, Jr. and D. P. Lathrop
{\it Phys. Rev. E}, {\bf 63}, 066211 (2001).

\bibitem{sweet2001b}
D. Sweet, E. Ott, T. M. Antonsen, Jr. and D. P. Lathrop, J. M. Finn
{\it Physics of Plasmas } {\bf 8}, 1944--1952 (2001).


\bibitem{phys-onoff}
    A. S. Pikovsky,{\it Z. Phys. B} {\bf 55}, 149 (1984);
    P.W. Hammer, N. Platt, S.M. Hammel, J.F. Heagy and B.D. Lee 
    {\it Phys. Rev. Lett.} {\bf 73} (8), 1095--1098 (1994).
    T. John, R. Stannarius and U. Behn, {\it Phys. Rev. Lett.} {\bf 83} (4), 749--752 (1999).
    D.L. Feng, C.X. Yu, J.L. Xie, W.X. Ding, {\it Phys. Rev. E} {\bf 58} (3), 3678--3685 (1998).
    F. Rodelsperger, A. Cenys and H. Benner, {\it Phys. Rev. Lett.} {\bf 75} (13), 2594--2597 (1995).

\bibitem{leprovost2006} N. Leprovost, B. Dubrulle, F. Plunian,
{\it Magnetohydrodynamics} {\bf  42} 131­--142 (2006).

\bibitem{ponty2007b}
Y. Ponty, J.-P. Laval , B. Dubrulle, F. Daviaud, J.-F. Pinton
{\it Phys. Rev. Lett.}, under press. arXiv:0707.2498

\bibitem{pinton2007} VKS Private communication, Les Houches, August 2007.


\bibitem{fujisaka}
    H. Fujisaka and T. Yamada, {\it Prog. Theor. Phys.} {\bf 74} (4), 918--921 (1984).
    H. Fujisaka, H. Ishii, M. Inoue and T.
    Yamada, {\it Prog. Theor. Phys.}  {\bf 76} (6), 1198--1209 (1986).

\bibitem{ott1004}  E. Ott and J. C. Sommerer,{\it Phys. Lett. A} {\bf 188}, 39 (1994).

\bibitem{Yu1990}   L. Yu, E. Ott, and Q. Chen, {\it Phys. Rev. Lett.} {\bf 65}, 2935 (1990).

\bibitem{Platt1994} N. Platt, S. M. Hammel, and J. F. Heagy, {\it Phys. Rev. Lett.} {\bf 72}, 3498 (1994).
\bibitem{Heagy1994} J. F. Heagy, N. Platt, and S. M. Hammel, {\it Phys. Rev. E} {\bf 49}, 1140 (1994).
\bibitem{Venka1995} S. C. Venkataramani, T. M. Antonsen, Jr., E. Ott, and J. C. Sommerer,
                    {\it Phys. Lett. A} {\bf 207}, 173 (1995) .
\bibitem{Venka1996} S. C. Venkataramani, T. M. Antonsen, Jr., E. Ott, and J. C. Sommerer,
                    {\it Physica D} {\bf 96}, 66 (1996).


\bibitem{aumaitre2005}
S. Auma\^itre, F. P\'etr\'elis, and K. Mallick,
{\it  Phys. Rev. Lett.} {\bf 95}, 064101 (2005).

\bibitem{aumaitre2006}
S. Auma\^itre, K. Mallick and F. P\'etr\'elis,
{\it Journal of Statistical Physics}, 2006, 123, 909--927 (2006).



\bibitem{cattaneo1996}
F. Cattaneo,  D.W. Hughes and E.J. Kim,
{\it Phys. Rev. Lett.} {\bf 76}, 2057­-2060 (1996).

\bibitem{zienicke1998}
 E. Zienicke, H. Politano and A. Pouquet,
{\it Phys. Rev. Lett.} {\bf 81}, 4640­-4640 (1998).

\bibitem{brummell} N.H. Brummell, F. Cattaneo , S.M. Tobias
{\it Fluid Dynamics Research} {\bf 28},  237­-265 (2001).



\bibitem{arnold1965} V. I. Arnold, {\it Comptes Rendus Acad. Sci. Paris} {\bf 261}, 17 (1965).

\bibitem{dombre1986} T. Dombre, U. Frisch, J. M. Greene, M. Henon, A. Mehr, and A. Soward,
{\it J. Fluid Mech.} {\bf 167}, 353 (1986).

\bibitem{galloway1992} Galloway, D.J. and Proctor, M.R.E.
{\it Nature} {\bf 356}, 691--693 (1992).

\bibitem{ponty1995} Y. Ponty, A. Pouquet and P.L. Sulem,
{\it Geophys. Astrophys. Fluid Dyn.} 79, 239­-257 (1995).

\bibitem{arnold1983} V.I. Arnold, and  E.I. Korkina,
{\it Vestn. Mosk. Univ. Mat. Mekh. } {\bf 3}, 43--46 (1983).

\bibitem{galloway1986} Galloway, D.J. and Frisch, U.,
{\it Geophys. Astrophys .Fluid Dyn.} {\bf 36}, 53--83 (1986).


\bibitem{galanti1992} B.Galanti, P. L. Sulem and A. Pouquet,
{\it Geophys. Astrophys .Fluid Dyn.} {\bf 66}, 183--208 (1992).

\bibitem{archontis2003} V. Archontis, S.B.F. Dorch and A. Nordlund,
{\it Astron. Astrophys.} {\bf 397}, 393­-399 (2003).

\bibitem{teyssier2006} R. Teyssier, S. Fromang and E Dormy
{\it J. Comp. Dyn.}  {\bf 218} 44--67 (2006).



\bibitem{childress1995}  S. Childress  and A.D. Gilbert,
``Stretch, Twist Fold: The Fast Dynamo'', {\it Springer-Verlag, New York} (1995).

\bibitem{moffat1985} H. K. Moffat and M. R. Proctor,
{\it J. Fluid Mech.} {\bf 154}, 493 (1985).

\bibitem{bayly1988}
B. J. Bayly and S. Childress,
{\it Geophys. Astrophys. Fluid Dyn.} {\bf 44}, 211 (1988).

\bibitem{finn1988a} J. M. Finn and E. Ott
{\it Phys. Fluids} {\bf 31}, 2992 (1988).

\bibitem{finn1988b}
John M. Finn and Edward Ott
{\it Phys. Rev. Lett.} {\bf 60}, 760 (1988).

\bibitem{podvigina1994} O.M. Podvigina and A. Pouquet,
{\it Physica D} {\bf 75}, 471­-508 (1994):

\bibitem{podvigina1999}
 O.M. Podvigina
{\it Physica D} {\bf 128},  250­-272 (1999).


\bibitem{mininni} 
P. D. Mininni 
{\it Physics of Plasmas} {\bf 13} (5), 056502 (2006).
  


\end{thebibliography}
\end{document}